\documentclass[A_4paper, 12pt, final]{article}
\pdfoutput=1

\usepackage{jheppub}

\usepackage[utf8]{inputenc}
\usepackage{graphicx}
\usepackage{hyperref}
\usepackage{amsmath}
\usepackage{color}
\usepackage{epstopdf}
\usepackage{float}
\usepackage{upgreek}
\usepackage{amssymb}
\usepackage{pdflscape}
\usepackage[nice]{nicefrac}
\usepackage[dvipsnames]{xcolor}
\usepackage{subcaption,graphicx}
\usepackage[numbers, sort&compress]{natbib}
\usepackage{braket}
\usepackage{comment}
\usepackage{slashed}
\usepackage{bbm}
\usepackage{commath}
\usepackage{mathtools}

\usepackage{xspace}
\newcommand*{\eq}{eq.\@\xspace}
\newcommand*{\cf}{cf.\@\xspace}
\newcommand*{\eqs}{eqs.\@\xspace}
\newcommand*{\eg}{e.g.\@\xspace}
\newcommand*{\ie}{i.e.\@\xspace}
\newcommand*{\fig}{fig.\@\xspace}

\newcommand*\diff{\mathrm{d}}
\newcommand*\ex{\mathrm{e}}

\title{Detecting Axion-Like Particles  With Coiled Optical Fibers I:
Silica Fibers}

\author[a,b]{Gia Dvali,}
\author[a,b]{Sebastian Zell,}
\author[a,b]{Tongxuan Zhang}

\affiliation[a]{Arnold Sommerfeld Center, Ludwig-Maximilians-Universit{\"a}t, Theresienstr.~37, 80333 M{\"u}nchen, Germany}
\affiliation[b]{Max-Planck-Institut f{\"u}r Physik, Boltzmannstr.~8, 85748 Garching, Germany}

\emailAdd{gdvali@mpp.mpg.de}
\emailAdd{sebastian.zell@lmu.de}
\emailAdd{Tongxuan.Zhang@lmu.de}

\date{}

\abstract{We propose a new approach to axion-like particle (ALP) searches based on long, coiled optical fibers in an external magnetic field. We develop the theoretical framework required to describe photon--ALP conversion in this geometry by incorporating transverse boundary conditions and fiber bending. For solid silica fibers with refractive index considerably larger than unity, we show that the leading signal is a phase shift of the photon, with negligible loss due to ALP production. This setup has the potential to set new constraints in the regime of large ALP mass. We further identify parameter regions in which boundary effects become important, in particular for hollow-core fibers, where signals due to ALPs can be significantly enhanced.}

\makeatletter
\gdef\@fpheader{\phantom{text}}
\makeatother

\begin{document}

\maketitle

\section{Introduction}

\subsection{Our proposal}

Axions
are among the most compelling candidates for physics beyond the Standard Model. Originally, 
they were introduced \cite{Weinberg:1977ma,Wilczek:1977pj}
in the context of the Peccei-Quinn solution \cite{Peccei:1977hh}
to the strong CP puzzle 
-- the striking experimental observation that quantum chromodynamics preserves CP symmetry to extremely high precision (see \cite{Abel:2020pzs} for the recent bound). In addition, axions are viable candidates for the microscopic origin of the dark matter observed in the Universe (see, e.g.,  \cite{Marsh:2015xka,DiLuzio:2020wdo,Choi:2020rgn} for reviews).  

However, the concept of axions extends well 
beyond the theory of the strong interactions. 
For clarity, following the usual convention, 
we shall refer to generalized axions as the \textit{axion-like particles} (ALPs). 
The defining features of such degrees of freedom are: {\it 1)} A pseudo-scalar (parity-odd) nature, 
{\it 2)}  periodicity, and {\it 3)} anomaly-type couplings to various dual gauge field strengths. 
The relation between their masses and couplings are model dependent and in what follows they  shall be treated as free parameters.

A broad experimental program is underway to detect axions and more generic ALPs
(see \eg \cite{PVLAS:2007wzd,ALPS:2009des,0910.5914,Irastorza:2011gs,CAST:2011rjr,Budker:2013hfa,Armengaud:2014gea,Kahn:2016aff,Brubaker:2016ktl,Caldwell:2016dcw,IAXO:2020wwp,DMRadio:2022pkf,Batllori:2023gwy,Batllori:2025hcl} and the overview in \cite{Adams:2022pbo}). These experiments exploit the universal interaction between the ALP field $a$ and electromagnetic fields,
\begin{equation}
	\label{fundamentalInteraction}
	\mathcal{L}_\text{int} = - g a \mathbf{E}\cdot \mathbf{B}\;,
\end{equation}
where $g$ denotes the photon--ALP coupling.
Here $\mathbf{E}$ and $\mathbf{B}$ stand for electric and magnetic components, respectively. In the presence of an external magnetic field, this interaction induces oscillations between photons and ALPs.

The probability for photon--ALP conversion increases with the distance over which the photon propagates through the magnetic field. Existing experimental designs achieve large effective interaction lengths in different ways, for instance by exploiting resonant enhancement in optical cavities \cite{ALPS:2009des, PVLAS:2007wzd, DellaValle:2014xoa, Betz:2013dza, 0910.5914, Brubaker:2016ktl, Ortiz:2020tgs, Batllori:2025hcl}. In this work, we propose a novel approach based on long, coiled optical fibers. Such fibers allow for very large propagation lengths while confining the magnetic field to a compact spatial region. Our proposal is inspired by the GRAVITES experiment \cite{Hilweg:2016oxz, ERC:GRAVITES}, which -- independently of ALPs -- employs a large-scale interferometer based on coiled fibers of length $L \sim 10^{5}\, \text{m}$ to measure gravitationally induced phase shifts of path-entangled photons.

So far, theoretical studies of photon--ALP conversion have been carried out in the plane-wave approximation \cite{Raffelt:1987im}. To describe propagation in bent fibers, these calculations must be extended in two essential ways. First, the transverse boundary conditions imposed by the fiber geometry must be taken into account. Second, the effects of bending must be incorporated consistently. Developing the corresponding theoretical framework is the primary goal of the present paper.

We focus initially on solid silica fibers with refractive index $n$ significantly larger than $1$. We show that, in this case, photon loss due to conversion into ALPs is negligible: ALPs produced through mixing remain effectively confined within the fiber, similarly to photons. Nevertheless, the presence of ALPs leads to a measurable effect. As we will demonstrate, the leading observable signal is a phase shift of the propagating photon mode, and ALPs can be detected provided
\begin{equation} \label{masterFormulaCoupling}
	g>\frac{1}{B}\sqrt{2n(n^2-1)\frac{\omega}{L}\delta \phi}\;. 
\end{equation}
Here $\omega$ is the photon frequency, $B$ the magnitude of the external magnetic field, $L$ the fiber length, and $\delta\phi$ the experimental sensitivity to phase shifts. Importantly, the sensitivity to the photon--ALP coupling improves with increasing fiber length.

As is evident from \eqref{masterFormulaCoupling}, the resulting bound on $g$ is independent of both the ALP mass $m_a$ and the transverse fiber size $l$. This behavior arises because the large refractive index induces an effective photon mass $m_\gamma^2 \equiv (1-n^2) \omega^2$, which dominates over all other relevant scales. This situation contrasts sharply with propagation in vacuum \cite{Raffelt:1987im}, where experimental sensitivity typically improves for smaller $m_a$. Consequently, our proposal is particularly well suited for probing ALPs with comparatively large masses. The precise lower bound on $m_a$ for which our setup can surpass existing terrestrial constraints depends on the achievable phase sensitivity $\delta\phi$. For example for $\delta \phi \sim 10^{-12}$, the sensitivity for measuring the coupling $g$ would become competitive for $m_a \gtrsim 10^{-2}\,\text{eV}$.

In the final part of the paper, we provide an outlook toward hollow-core fibers, where the refractive index $n$ is close to unity. In this case, parameters can be chosen such that ALPs are no longer confined and moreover resonant photon--ALP conversion becomes possible. We therefore expect hollow-core fibers to offer substantially enhanced sensitivity compared to solid silica fibers with the potential to probe a much larger region of parameter space. Indeed, the use of straight hollow-core fibers to search for light weakly interacting particles such as ALPs has already been proposed in the WISPFI experiment \cite{Batllori:2023gwy,Batllori:2025ogo}. However, in estimating the expected sensitivity, the role of transverse boundary conditions and the associated length scale $l$ was neglected in \cite{Batllori:2023gwy,Batllori:2025ogo}. We will show that this scale plays a crucial role in hollow-core fibers, implying that the corresponding sensitivity estimates must be revisited.

\subsection{Theoretical motivation}

Before proceeding with the experimental proposal, we shall discuss some theoretical motivations for searching for ALPs with masses $m_a \gtrsim 10^{-2}\,\text{eV}$. It is often argued that in this parameter space the strongest constraints on $g$ come from astrophysical observations, in particular from stellar production. However, such bounds  rely on assumptions about ALP interactions at high energies, for instance in stellar interiors.   This is certainly applicable if the coupling  can be safely extrapolated all the way to energies comparable to the star temperature. 
 However, the validity of such extrapolation cannot be guaranteed without actual knowledge of ALP's effective theory at the corresponding scale.  
 In the present analysis we shall take a more open-minded attitude and
  relax any assumption about the knowledge of ALP's behaviour much above the energy scale relevant for the considered experimental setup. 
As a very crude guideline, we shall mention some motivating proposals for ALPs to which stellar bounds do not apply.

In particular, ALP production may be suppressed at high energies if ALPs arise as composite states of Standard Model particles such as neutrinos \cite{Dvali:2013cpa, Dvali:2016uhn, Dvali:2016eay, Dvali:2021uvk}.  In this picture, above the compositeness scale, the ALP is no longer an independent  degree of freedom and ``melts'' into ordinary neutrinos.\footnote{Another interesting candidate for a light composite ALP that is not constrained by astrophysical processes is the recently-proposed  $\eta_{\rm w}$-meson originating from the 
$B+L$-symmetry-violating condensate of the fermionic  't Hooft determinant of weak-doublet quarks and leptons \cite{Dvali:2024zpc, Dvali:2025pcx}. This condensate is generated by the electroweak instantons.  However, the origin of the $\eta_{\rm w}$'s coupling to photons (\ref{fundamentalInteraction}), that would be relevant for the presented search method, is currently unclear.} Correspondingly, the bounds from the star cooling are abolished.

 In this proposal the compositeness of ALPs 
 originates from non-perturbative gravitational effects and is constrained to be around the neutrino mass $m_{\nu}$. In fact, in \cite{Dvali:2016uhn} this effect was suggested to be the origin of the neutrino masses. This idea is supported by a striking coincidence that $m_{\nu}$ is of the same order of magnitude as the scale beyond which modifications of ordinary gravity are  experimentally unconstrained. The current bound on this scale, $30\,\mu\text{m} \sim (10^{-2}\,{\rm eV})^{-1}$, comes from the 
 measurements of Newtonian gravity 
\cite{Adelberger:2009zz, Tan:2016vwu, Lee:2020zjt}.
 Previously  \cite{Arkani-Hamed:1998wuz, Dvali:1999cn}, an explanation of this coincidence was provided from the gravitational origin of small $m_{\nu}$ in the framework of large extra dimensions \cite{Arkani-Hamed:1998jmv, Antoniadis:1998ig, Arkani-Hamed:1998sfv}. 

 However, for the current discussion more relevant is the alternative proposal of the gravitational origin of the neutrino mass \cite{Dvali:2016uhn}, which is  accompanied by the emergence of a composite ALP with the mass around 
 $m_{\nu}$. Rather nicely, the suggested experimental setup of optical fibers offers access precisely to this 
 energy scale.  
  In this context, we can argue that a signature of ALP in our proposed experiment is expected to be 
correlated with deviations from Newtonian gravity at scales of $\sim \mu\text{m}$.
This adds motivation to future improved-precision measurements of short-distance gravity.\footnote{We thank Markus Aspelmeyer for communication about the future prospects of short-distance Newtonian gravity measurements.}

We note that a suppression 
 of the effective ALP coupling $g$ at high energies could also occur for ALPs exhibiting the phenomenon of so-called ``infrared transparency'', originally discussed in \cite{Dvali:2001gx} in the context of a high dimensional gravity model \cite{Dvali:2000hr}.  Similarly to the case of compositeness,  the interaction softens above a certain critical scale and the UV-modes effectively decouple.\footnote{In terms of effective theory the effect would amount to the effective ALP-photon coupling,
$g(p)$, becoming a form-factor that diminishes for processes with momentum transfer above the critical scale.}
  Again, we must stress that in the present context we refer to this effect solely as a very general motivating factor without the goal of entering into any realistic model building.
  
Overall, from the point of view of fundamental physics particularly 
interesting targets are theories  in which in UV the effective coupling of ALPs to ordinary matter is strongly reduced, significantly weakening or eliminating astrophysical bounds. In such scenarios, laboratory-based experiments become the primary avenue for discovery, making terrestrial production schemes -- such as the one proposed here -- especially relevant.  

The organisation of the paper is as follows: We shall restrict ourselves to plane waves in section \ref{sec:planeWave} and show how \eq \eqref{masterFormulaCoupling} can be derived in this approximation. What remains to do is to justify that confinement and bending can indeed be neglected, which we achieve in section \ref{sec:ConfinementBending}. In section \ref{sec:summary}, we shall derive the control parameter that determines whether ALPs are confined to the fiber and use it to give an outlook to hollow-core fibers, showing that the transverse scale $l$ plays an important role in this case and that much stronger signals can be achieved. We conclude in section \ref{sec:conclusion}, and appendices \ref{app:generalSolution} and \ref{app:photonInBentWaveguide} contain details about initial conditions for propagation and the photon solution in the bent waveguide, respectively, while appendix \ref{app:HamiltonianMixing} provides a complementary understanding of our findings in terms of Hamiltonian mixing.

\section{Photon--ALP conversion}
\label{sec:planeWave}

\subsection{The model}

The following analysis of the photon--ALP conversion is based on \cite{Raffelt:1987im}, but we use the electric field strength as the photon degree of freedom rather than the potential.
The full Lagrangian describing the interaction between the photon and ALP fields in vacuum reads\footnote{We work in natural units $c=1/\sqrt{\mu_0\epsilon_0}=1$, where $\epsilon_0$ and $\mu_0$ are the vacuum permittivity and permeability, respectively. We further choose Heaviside-Lorentz units commonly used in particle physics, in which $\epsilon_0=\mu_0=1$, such that no irrational factors appear in the Lagrangian (in contrast to $\mathcal{L}\equiv-\frac{1}{4\mu_0}F_{\mu\nu}F^{\mu\nu}=-\frac{1}{16\pi}F_{\mu\nu}F^{\mu\nu}$ in Gaussian units where $\epsilon_0^{-1}=\mu_0=4\pi$).}
\begin{equation}
    \mathcal{L}=-\frac{1}{4}F_{\mu\nu}F^{\mu\nu}+\frac{1}{2}(\partial_\mu a)^2 -\frac{1}{2}m_a^2 a^2+\frac{1}{4}g a F_{\mu\nu}\Tilde{F}^{\mu\nu}\;,
\end{equation}
 where as before $m_a$ is the mass of the ALP and $g$ sets the strength of the photon--ALP coupling. Moreover, $F_{\mu\nu}=\partial_\mu A_\nu-\partial_\nu A_\mu$ is the electromagnetic tensor and $\Tilde{F}^{\mu\nu}=\frac{1}{2}\varepsilon^{\mu\nu\sigma\rho}F_{\sigma\rho}$ represents its dual, with $\varepsilon^{0123}=1$. As is well-known,  $F_{\mu\nu}\Tilde{F}^{\mu\nu} = - 4 \mathbf{E}\cdot \mathbf{B}$ and so the interaction term coincides with the one shown in \eq \eqref{fundamentalInteraction}.

We can generalize the above system to an isotropic, homogeneous, and linear background medium with constant relative permittivity $\epsilon_r$ and relative permeability $\mu_r$. The Lagrangian reads (see \cite{Brevik:2022gkt, Millar:2016cjp}) 
\begin{equation}
    \begin{aligned}
        \mathcal{L}&=-\frac{{1}}{4}F_{\mu\nu}D^{\mu\nu}+\frac{1}{2}(\partial_\mu a)^2-\frac{1}{2}m_a^2a^2+\frac{1}{4}g aF_{\mu\nu}\tilde{F}^{\mu\nu}\\
        &=\frac{{1}}{2}({\epsilon_r}\mathbf{E}^2-\frac{1}{{\mu_r}}\mathbf{B}^2)+\frac{1}{2}(\partial_\mu a)^2-\frac{1}{2}m_a^2a^2-g a\mathbf{E}\cdot\mathbf{B}\;,
    \end{aligned}
\end{equation}
where $D^{\mu\nu}$ is the electromagnetic displacement tensor, 
\begin{equation}
    D^{\mu\nu}=\begin{pmatrix}
        0 & -D_x & -D_y & -D_z \\
        D_x & 0 & -H_z & H_y \\
        D_y & H_z & 0 & -H_x \\
        D_z & -H_y & H_x & 0
    \end{pmatrix}\;, 
\end{equation}
and its elements fulfill the linear relations $\mathbf{D}={\epsilon_r}\mathbf{E},\ \mathbf{H}=\frac{1}{{\mu_r}}\mathbf{B}$. 

Now we add a strong external magnetic field $\mathbf{B}_{\text{ext}}=B\hat{\mathbf{x}}$ perpendicular to the propagation direction $z$, whose strength is much larger than that of the propagating modes such that we can approximate $\mathbf{B}\approx \mathbf{B}_\text{ext}$ in the interaction term. Assuming the incident photon field propagating along the $z$-axis is uniform in the transverse directions, one can derive the equations of motion for $E_x$ and $a$ as a $1+1$-dimensional problem
\begin{equation} \label{eomPlane}
    \begin{aligned}
        (n^2\partial_t^2-\partial_z^2)E_x&={\mu_r}gB\partial_t^2a\;,\\
        (\partial_t^2-\partial_z^2+m_a^2)a&=-gBE_x\;,  
    \end{aligned}
\end{equation}
where the refractive index $n=\sqrt{\mu_r\epsilon_r}$ is a constant. Although the relative magnetic permeability affects the mixing term in the first equation, the influence is negligible since $\mu_r\approx 1$ for the common fibers made of silica,
and so we set $\mu_r=1$ in what follows.\footnote
{If we keep $\mu_r$ in the calculation, the result will be modified by taking $\omega\to\sqrt{\mu_r}\omega$. It is still interesting to consider the medium with a large $\mu_r$ in the strong external magnetic field. }

For now we shall assume time harmonic solutions, $E_x \sim e^{-i\omega t}$ and $a\sim e^{-i\omega t}$, where $\omega$ denotes the energy of the incident photon. Then the equations of motion can be collected in a matrix form
\begin{equation}\label{eomGeneral}
	\left[\partial_z^2+\omega^2+\begin{pmatrix}
		-m_\gamma^2 & gB \omega \\
		gB \omega & -m_a^2
	\end{pmatrix}\right]\begin{pmatrix}
		\gamma \\ a
	\end{pmatrix}=0\;,
\end{equation}
where we defined $\gamma\equiv -E_x/\omega$ and recall that $m_\gamma^2=(1-n^2)\omega^2<0$ for normal silica fibers.

\subsection{General solution}
To solve \eq \eqref{eomGeneral}, we define the diagonal fields
\begin{equation}
\begin{pmatrix}
	\gamma' \\ a'
\end{pmatrix} =	V \begin{pmatrix}
		\gamma \\ a
	\end{pmatrix}\;, \qquad V = \begin{pmatrix}
	\cos\theta & \sin\theta \\ -\sin\theta & \cos\theta
	\end{pmatrix} \;.
\end{equation}
Then the mixing angle $\theta$ fulfills
\begin{equation}\label{mixingAngleGeneral}
	\frac{1}{2}\tan2\theta=\frac{gB\omega}{m_a^2-m_\gamma^2}\;,
\end{equation}
and yields
\begin{equation}\label{eomGeneralDiagonal}
	\left[\partial_z^2+\omega^2+\begin{pmatrix}
		\lambda_+^2 & 0 \\ 0 & \lambda_-^2
	\end{pmatrix}\right]\begin{pmatrix}
		\gamma'\\ a'
	\end{pmatrix}=0\;, 
\end{equation}
where 
\begin{align}
\lambda_+^2  & = g B  \omega  \sin 2 \theta -m_a^2 \sin^2\theta -m_\gamma^2 \cos ^2\theta \;, \\
\lambda_-^2 & = -g B  \omega  \sin 2\theta -m_a^2 \cos ^2\theta -m_\gamma^2 \sin^2\theta\;.
\end{align}
If pure photons are incident, \ie
\begin{equation} \label{initialConditionsPureIncident}
	\begin{pmatrix}
		\gamma \\ a
	\end{pmatrix}_\text{in}=\begin{pmatrix}
		\mathcal{A} \\ 0
	\end{pmatrix}\;,
\end{equation}
a particular oscillating solution of \eq\eqref{eomGeneral} can be constructed as 
\begin{equation}\label{solGeneral}
	\begin{pmatrix}
		\gamma(z) \\ {a}(z) 
	\end{pmatrix}=V^{-1}\begin{pmatrix}
		\ex^{ik_+'z} & 0 \\
		0 & \ex^{ik_-'z}
	\end{pmatrix}V\begin{pmatrix}
		\gamma \\ a
	\end{pmatrix}_\text{in}= \mathcal{A}
	\begin{pmatrix}
		\cos^2 \theta \ex^{ik_+'z}+\sin^2\theta \ex^{ik_-'z} \\
		\cos\theta\sin\theta(\ex^{ik_+'z}-\ex^{ik_-'z})
	\end{pmatrix}\;, 
\end{equation}
where the momenta are defined by
\begin{equation} \label{momenta}
	k'_\pm=\sqrt{\omega^2+\lambda^2_\pm}\;.
\end{equation}
Importantly, the initial conditions \eqref{initialConditionsPureIncident} do not specify the derivative of the fields, and so they are not sufficient to single out a unique solution. In appendix \ref{app:generalSolution}, we show the most general solution and how it reduces to the above purely incident photon case. 

For $gB\omega \ll m_a^2-m_\gamma^2$, we can approximate to second order in small $\theta$: 
\begin{align}
	\theta &\approx\frac{gB\omega}{m_a^2-m_\gamma^2} \;,\\
	\lambda_+^2 & \approx -m_\gamma^2 + \theta^2   \left(m_a^2-m_\gamma^2\right) \;,\\
	\lambda_-^2 & \approx -m_a^2 - \theta^2   \left(m_a^2-m_\gamma^2\right)\;.
\end{align}
Then the momenta \eqref{momenta} become
\begin{align}
	k'_+ & \approx \sqrt{\omega^2 - m_\gamma^2}\left(1 + \frac{1}{2} \theta^2 \frac{m_a^2-m_\gamma^2}{\omega^2 - m_\gamma^2}\right) \;,\\
	k'_- & \approx \sqrt{\omega^2 - m_a^2} \left(1 - \frac{1}{2} \theta^2 \frac{m_a^2-m_\gamma^2}{\omega^2 - m_a^2}\right) \;,
\end{align}
and the oscillating solution \eqref{solGeneral} gives after a distance $L$
\begin{equation} \label{solGeneralLeading}
	\begin{pmatrix}
		\gamma(L) \\ {a}(L) 
	\end{pmatrix}\approx \mathcal{A} 	\ex^{i\sqrt{\omega^2 - m_\gamma^2} L}\begin{pmatrix}
	1-\rho_\gamma(L)+i\phi_\gamma(L) \\
	\rho_a(L)+i\phi_a(L)
	\end{pmatrix}\;,
\end{equation}
where
\begin{align}
		&\rho_\gamma(L)=2\theta^2\sin^2\left(\frac{\Delta \omega L}{2}\right)\;,
		&\phi_\gamma(L)&=\theta^2\left(\frac{\left(m_a^2-m_\gamma^2\right)L}{2\sqrt{\omega^2-m_\gamma^2}}-\sin \Delta\omega L\right)\;, \label{masterPhoton}\\
		&\rho_a(L)=2\theta\sin^2\left(\frac{\Delta \omega L}{2}\right)\;,
		&\phi_a(L)&=\theta\sin \Delta \omega L\;, \label{masterALP}
\end{align}
and we defined
\begin{equation}
	\Delta \omega = \sqrt{\omega^2 - m_\gamma^2} - \sqrt{\omega^2 - m_a^2} \;.
\end{equation}

\subsection{Different setups}

\subsubsection{Plane wave in vacuum}

For a plane wave in vacuum, $m_\gamma=0$ and so
\begin{equation}
	\theta =\frac{gB\omega}{m_a^2} \;, \qquad 	\Delta \omega \approx \frac{m_a^2}{2 \omega} \;,
\end{equation}
where we also assumed $m_a \ll \omega$. Then \eqs \eqref{masterPhoton} and \eqref{masterALP} give
\begin{equation}\label{resultVacuum}
	\begin{aligned}
		&\rho_\gamma(L)=2\theta^2\sin^2\left(\frac{m_a^2 L}{4\omega}\right)\;,
		&\phi_\gamma(L)&=\theta^2\left(\frac{m_a^2L}{2\omega}-\sin\frac{m_a^2 L}{2\omega}\right)\;,\\
		&\rho_a(L)=2\theta\sin^2\left(\frac{m_a^2L}{4\omega}\right)\;,
		&\phi_a(L)&=\theta\sin\frac{m_a^2L}{2\omega}\;,
	\end{aligned}
\end{equation}
in accordance with \cite{Raffelt:1987im}.
We note that the argument in the oscillating sine functions takes the above form only when the distance $L$ is not large such that the higher order terms, $O((m_a^2/\omega^2)^2)\omega L$, can be neglected. Otherwise, we should keep $\Delta\omega L$ for correctness, though it will not influence the detection due to the high oscillation frequency. 

Purely terrestrial experiments, which aim to detect ALPs without assuming that they exist in the Universe or in our surroundings, can be classified into two types according to the different quantities in \eq\eqref{resultVacuum} they measure. The first type corresponds to ``Light-Shining-Through-Walls (LSW)'' experiments, in which a strong magnetic field converts photons to ALPs, the ALPs then travel through a wall that is impassable to photons, and finally another strong external magnetic field is used behind the wall to convert the passed ALPs into photons as the ``shining'' we can detect. Therefore, the quantity to be measured in this experiment is the twice-conversion probability\footnote
{Notice that $P_{\gamma\to a}=P_{a\to \gamma}$ because of the symmetry.}
\begin{equation}
    P_{\gamma\to a}^2(L)=\left(\abs{\rho_a+i\phi_a}^2\right)^2=\left[4\theta^2\sin^2\left(\frac{m_a^2 L}{4\omega}\right)\right]^2 \leq 16\frac{(gB\omega)^4}{m_a^8}\;. 
\end{equation}

\begin{figure}
	\centering
	\includegraphics[width=0.9\linewidth]{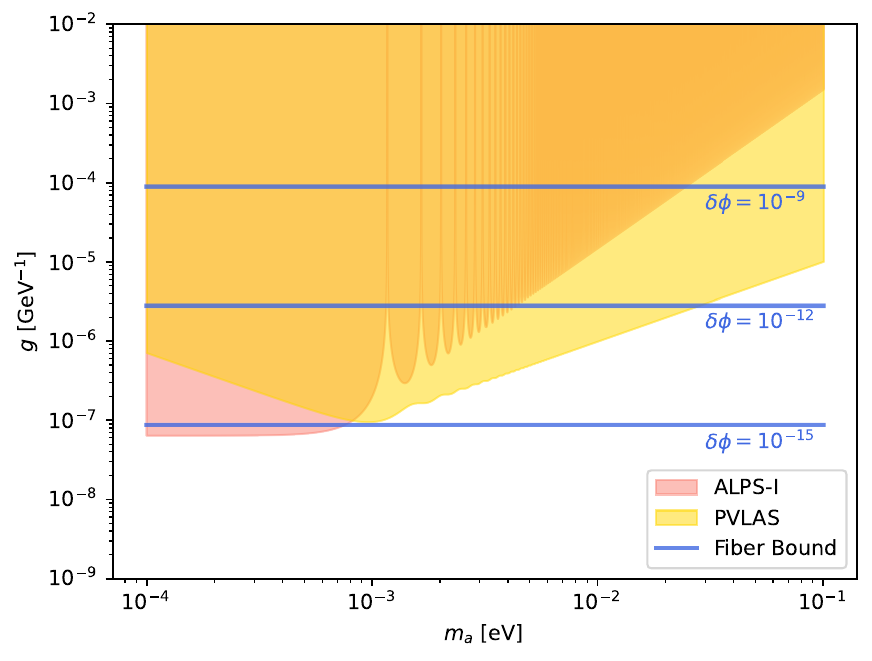}
	\caption{The exclusion plot for previous experiments and the proposed fiber-based interferometer. The results shown for comparison are taken from ALPS-I \cite{Ehret:2010mh} and PVLAS \cite{PVLAS:2007wzd} experiments. The parameters used for the calculation of the fiber bound according to \eq\eqref{masterFormulaCoupling} are $n=1.5,\,\omega=1\,\mathrm{eV},\,B=5\,\mathrm{T},\,L=10^5\,\mathrm{m},$ and different values of $\delta\phi$ shown in the plot below each bound.}
		\label{fig: exclusion plot}
	\end{figure}

The second type of measurement refers to photon birefringence effects. For a linearly polarized laser beam as the incident state, the final state after traveling the distance $L$ will be elliptically polarized with a rotation of the major axis. The ellipticity and the rotation angle are determined by $\phi_\gamma(L)$ and $\rho_\gamma(L)$, respectively. The latter signal is bounded by 
\begin{equation}
    \rho_\gamma(L)=2\left(\frac{gB\omega}{m_a^2}\right)^2\sin^2\left(\frac{m_a^2L}{4\omega}\right)\leq 2\frac{(gB\omega)^2}{m_a^4}\;. 
\end{equation}
Therefore, the sensitivity for measuring the transition probability or the rotation angle at least decays as $m_a^{-4}$ for large $m_a$. In contrast, the phase shift $\phi_\gamma(L)$ is dominated by a term that grows linearly with $L$:  
\begin{equation}
    \phi_\gamma(L)=\left(\frac{gB\omega}{m_a^2}\right)^2\left(\frac{m_a^2L}{2\omega}-\sin\frac{m_a^2 L}{2\omega}\right) \approx \frac{(gB)^2\omega L}{2m_a^2}\;, 
\end{equation}
which shows that sensitivity only decreases proportionally to $m_a^{-2}$. Thus, the latter would be a better choice for detecting photon--ALP conversion in the large ALP mass range, where a longer travel distance $L$ leads to a more significant signal. For completeness, as $m_a$ goes small, the oscillation terms would be important, and the transition occurs at
\begin{equation}
    m_a\sim \sqrt{\frac{\omega}{L}}\sim 10^{-3}\text{eV}\;,  
\end{equation}
where we take the typical values $L\sim 1\,\text{m}$ and $\omega\sim1\,\text{eV}$ used in the previous experiments. A sketch of exclusion regions from ALPS-I \cite{Ehret:2010mh} and PVLAS \cite{PVLAS:2007wzd} is shown in \fig\ref{fig: exclusion plot}. 

Apart from the direct detection of the birefringence, the amplitude change and the phase shift of incident photons can also be measured via interferometry. By putting one arm of the interferometer into a strong external magnetic field while leaving the other arm free, the birefringent information will be encoded in its fringes. The use of Michelson interferometers to detect ALPs was proposed in \cite{Tam:2011kw}. And before that, large-scale gravitational wave interferometers for observing nonlinear QED effects, which is closely related to ALP detection, were suggested by \cite{Boer:2002zw}. Although the scale of the interferometer can be large, it is difficult to achieve a long travel distance $L$ with the extent of strong magnetic fields. However, the large-scale fiber-based interferometer has the potential to keep the scale of the traveling distance in the magnetic field still large by wrapping the long fiber as a coil, which can be put into the whole magnetic region. Then, it is suitable for detecting ALPs with a large mass. In contrast to the photon--ALP conversion in vacuum, the optical fiber provides a different background for photons traveling due to its material and geometry. We start with the discussion of the large-refractive-index medium for ordinary fibers in the next section. 

\subsubsection{Plane wave in medium}

For a plane wave in a medium, we get
\begin{equation}
	\theta \approx\frac{gB\omega}{-m_\gamma^2}=\frac{gB}{(n^2-1)\omega} \;, \qquad 	\Delta \omega =n \omega -\sqrt{\omega^2 - m_a^2} \;,
\end{equation}
where we assumed $|m_\gamma^2| \gg m_a^2$ in the first equation. Then \eqs \eqref{masterPhoton} and \eqref{masterALP} give
\begin{equation}\label{resultMedium}
	\begin{aligned}
		&\rho_\gamma(L)=2\theta^2\sin^2\left(\frac{\Delta\omega L}{2}\right)\;,
		&\phi_\gamma(L)&=\theta^2\left(\frac{(n^2-1)\omega L}{2n}-\sin\Delta\omega L\right)\;,\\
		&\rho_a(L)=2\theta\sin^2\left(\frac{\Delta\omega L}{2}\right)\;,
		&\phi_a(L)&=\theta\sin\Delta\omega L\;.
	\end{aligned}
\end{equation}
The signal relevant for detection comes from the phase shift of the photon field:
\begin{equation} \label{masterFormulaPhase}
	\phi_\gamma(L)\approx \frac{(gB)^2}{(n^2-1)^2\omega^2}\left(\frac{n^2-1}{2n}\omega L\right)\;, 
\end{equation}
where we dropped the oscillation term. If $\delta \phi$ is the experimental sensitivity for measuring the phase, then we can detect the signal from the photon--ALP coupling if $\phi_\gamma(L)>\delta \phi$, which gives \eq \eqref{masterFormulaCoupling} shown in the introduction.

To estimate the bound, we can pick the approximate parameters $n=1.5$, $\omega=1\,\text{eV}$, the long distance $L=10^5\,\text{m}$ used in GRAVITES \cite{Hilweg:2016oxz, ERC:GRAVITES}, and the external $B=5\,\text{T}\sim 10^3\,\text{eV}^2$ as was taken in ALPS-I \cite{Ehret:2010mh}. For these values, we show the expected sensitivity in \fig \ref{fig: exclusion plot} and compare it to previous laboratory experiments. We see that $\delta\phi=10^{-12}$ would already set new constraints in the large mass region.\footnote
{In GRAVITES, the phase sensitivity $4.42\times 10^{-6}$ rad root-mean-square (RMS) was reached in a 50-km fiber interferometer operating at a single photon level \cite{Yu:2025cch}. We expect that the sensitivity can be improved with higher photon intensities. On the optimistic side, the proposed WISPFI experiment \cite{Batllori:2023gwy} reports an anticipated detector-noise-limited sensitivity of $\sim 10^{-18}$ to the relative amplitude change (not taking into account additional experimental noise sources), with a mechanism that can potentially also be applied to interferometric phase-shift detection. 
}

\subsubsection{Propagation in a waveguide}

So far, we have assumed that the photon and ALP fields are uniform in the transverse directions and that the propagation occurs along a straight path. Evidently, this is not the case in a fiber, and so it is unclear if \eq \eqref{masterFormulaCoupling} applies to a realistic fiber. In particular, one may naively expect that a part of ALPs leak from the waveguide after they are produced such that the amplitude of photon field will decrease with $L$. In the following section, we will study the effects of the waveguide in detail. We find that the transverse boundary conditions and the bending only lead  to small corrections, with ALP leakage being negligible. Therefore, our formula \eqref{masterFormulaCoupling} is applicable to normal silica waveguides. 

Of course, realistic implementations must take into account various noise effects that we have neglected. These arise from imperfections in the fiber material and geometry, as well as from environmental and thermal perturbations, which can modify the fiber length, refractive index, or birefringence and excite acoustic modes (see the overview in \cite{Hilweg:2022ubt}). The application of an external magnetic field may introduce additional noise sources, for example through the Faraday effect and temporal or spatial variations of the magnetic field.

\section{Effects of transverse boundary conditions and bending}
\label{sec:ConfinementBending}

\subsection{Confinement in straight waveguides}

To allow for simple analytic calculation, we shall consider an over-simplified model of a waveguide. In the core, we consider a dielectric medium with refractive index $n>1$, as is the case for a normal silica fiber. Instead of the cladding, however, we shall assume that the boundary is made of a perfect conductor. In other words, we are using a medium-filled metallic waveguide. Thus, while a realistic fiber would have evanescent waves outside the core, we impose the boundary condition $E_t=0$ at the interface, where $E_t$ denotes the tangent component of the electric field. We note that this interface condition does not change even if the photon--ALP conversion is introduced since it is derived from the identity $\nabla \times \mathbf{E}=-\frac{\partial \mathbf{B}}{\partial t}$. Moreover, we take the boundary to be rectangular. As we shall see, qualitative conclusions only depend on the compact transverse photon profile and on the relationship of $\omega$, $\beta$ and $m_a$, and therefore are independent of the detailed boundary model.

 Without the conversion to ALPs, the solution for the $x$-component electric field reads
\begin{equation}\label{Ex pure photon}
    \begin{aligned}
        E_x(t,x,y,z)&=E_x(x,y,z)e^{-i\omega t}\\
        &=E_0\cos(k_x x)\sin{(k_y y)}\mathbf{1}_{\{0\leq x\leq l_x\}}\mathbf{1}_{\{0\leq y\leq l_y\}}e^{-i\omega t+i\beta z}\;,
    \end{aligned}
\end{equation}
where \(k_x l_x=m_x \pi, \, k_y l_y=m_y \pi \; (m_{x,y}\in \mathbb{Z})\), \(l_x, l_y\) denote the size of the rectangle, and $\mathbf{1}_{\{\dots\}}$ the indicator function. The longitudinal momentum $\beta$ is determined by the dispersion relation
\begin{equation}
    n^2\omega^2=\beta^2+k_x^2+k_y^2\;.
\end{equation}
For simplicity, we pick the mode \(m_x=0, m_y=1\) such that $E_x$ becomes uniform along the \(x\)-direction. It is convenient to rewrite $k_y$ as $k_\perp$, $l_y$ as $l$, and the dispersion relation as $\omega^2=\beta^2+k_\perp^2+m_\gamma^2$. As a consequence, the ALP field is also $x$-independent and the conversion problem becomes $2+1$ dimensional. The whole system is described by the following equations of motion in the core of the waveguide
\begin{equation}\label{eomGuided}
    \begin{aligned}
        (n^2\partial_t^2-\partial_y^2-\partial_z^2)E_x(t,y,z)&=gB\partial_t^2 a(t,y,z)\;,\\
        (\partial_t^2-\partial_y^2-\partial_z^2+m_a^2)a(t,y,z)&=-gB E_x(t,y,z) \;,
    \end{aligned}
\end{equation}
to be compared with \eq \eqref{eomPlane}.

\subsubsection{Direct solution and energy analysis}
\label{sssec:energyAnalysis}

For a small coupling constant $g$, we can solve the system perturbatively. Namely, we take the pure photon solution \eqref{Ex pure photon} as the source in the equation of the ALP field. Once the ALP field $a\sim g$ is solved, we then plug it into the first equation to study the correction $\sim g^2$ to the photon field. With a same field redefinition as before, $\gamma\equiv -E_x/\omega$, we seek a solution to 
\begin{equation}
    \begin{aligned}
        &(\partial_t^2-\partial_y^2-\partial_z^2+m_a^2)a=\theta(m_a^2-m_\gamma^2)\gamma^{(0)}\;, 
    \end{aligned}
\end{equation}
where
\begin{equation}\label{photonConfined}
    \gamma^{(0)} = \mathcal{A} \sin\left(k_\perp y\right) \mathbf{1}_{\{0\leq y\leq l\}}\ex^{-i\omega t + i\beta z}\;, \qquad \theta=\frac{gB\omega}{m_a^2-m_\gamma^2}\;. 
\end{equation}
We can split the ALP solution as
\begin{equation} \label{ALPSplit}
	a = \theta\left(a_\gamma + a_{\text{free}}\right) \;,
\end{equation}
where
$a_{\text{free}}$ fulfills the free wave equation
\begin{equation} \label{ALPFree}
	\left(\partial_t^2 - \partial_y^2 - \partial_z^2 + m_a^2\right)	a_{\text{free}} = 0 \;,
\end{equation}
and the photon-induced part $a_\gamma$ is given by,
\begin{align}
a_\gamma	&= \mathcal{A}\ex^{-i\omega t + i\beta z} f_a(y) \;, \label{aGamma}\\
	f_a(y) &= \begin{cases}\frac{k_\perp}{2 |\mu|} \left(\ex^{|\mu| y}+\ex^{|\mu| (y-l)}\right) & y<0\\
		\sin\left(k_\perp y\right) + \frac{k_\perp}{2 |\mu|} \left(\ex^{-|\mu| y}+\ex^{|\mu| (y-l)}\right)& 0\leq y \leq l \\
		\frac{k_\perp}{2 |\mu|} \left(\ex^{-|\mu| y}+\ex^{-|\mu| (y-l)}\right) & y>l 
	\end{cases} \;, \label{fa}
\end{align}
and we expressed $f_a(y)$ in terms of 
\begin{equation} \label{musquare}
\mu^2 \equiv \omega^2 - \beta^2 - m_a^2 \;.
\end{equation}
The fact that $\mu^2<0$ determines the structure of the solution -- for positive $\mu^2$ the argument of the exponentials would be imaginary, and so we would obtain ALP waves propagating away from the fiber instead of the evanescent functions of \eq \eqref{fa}.

Imposing the boundary condition of no ALPs in the initial state gives
\begin{align} \label{initialConditionsALPs}
	& a(t=0,y,z)=0 \;, && \partial_t a(t=0,y,z) = 0\;,\\
	\Leftrightarrow \quad   & a_{\text{free}}(t=0,y,z)=-\mathcal{A}\ex^{ i\beta z} f_a(y) \;, && \partial_t a_{\text{free}}(t=0,y,z) = i\mathcal{A} \omega \ex^{ i\beta z} f_a(y)  \;.
\end{align}
We now compute the energy:
\begin{equation} \label{energyAverage}
    \begin{aligned}
        E &=\frac{1}{2}\int\diff x\, \diff y\, \diff z\left((\Re(\partial_ta))^2+(\Re(\nabla a))^2+m_a^2(\Re(a))^2 \right)\\
        &=\frac{1}{4}l_x L\int\diff y\, \Re\left(\partial_t a\partial_t a^*+\nabla a\cdot\nabla a^*+m_a^2 a a^*\right)\equiv\frac{1}{4}l_xL(\mathcal{E}_0+\mathcal{E}_t)\;,
    \end{aligned}
\end{equation}
where
\begin{equation}
		\mathcal{E}_0  = 2\theta^2\mathcal{A}^2\int \diff y \left[\left(2\omega^2 - \mu^2 \right) f_a(y)^2 + f_a'(y)^2 \right] \;,
\end{equation}
and 
\begin{align}
		\mathcal{E}_t &= 2\theta^2\int \diff y\, \Re\left(-i\omega a_\gamma \partial_t a_{\text{free}}^* +i\beta a_\gamma \partial_z a_{\text{free}}^*  +\partial_y a_\gamma \partial_y a_{\text{free}}^*+ m_a^2 a_\gamma a_{\text{free}}^* \right) \;.
\end{align}
Here $\Re$ denotes the real part and in the second line of \eq \eqref{energyAverage} we implicitly averaged over the longitudinal direction $z$. The term $\mathcal{E}_0$ comes from two identical contributions quadratic in $a_\gamma$ and in $a_{\text{free}}$, respectively, and in evaluating the latter we used that \eq \eqref{ALPFree} conserves energy and so we can use the initial conditions \eqref{initialConditionsALPs} at $t=0$. Thus, $\mathcal{E}_0$ is time-independent, and it is easy to check that $\mathcal{E}_t (t=0) = - \mathcal{E}_0$.

Using \eq \eqref{fa}, we can explicitly compute $\mathcal{E}_0$. Clearly, the answer is finite and for large $|m_\gamma|$, we get $\mathcal{E}_0 \sim \theta^2\mathcal{A}^2\omega^2 l$. Without knowing $a_{\text{free}}(t)$, we cannot compute $\mathcal{E}_t (t)$, but we can bound it. It suffices to know that $a_{\text{free}}(t)$ is bounded, $|a_{\text{free}}(t)|\lesssim 1$ (with the same property holding for its derivatives), and that $a_\gamma$ (and its derivatives) decay exponentially outside the fiber. Therefore, $\mathcal{E}_t$ is bounded for all times $t$ with $|\mathcal{E}_t|\lesssim \theta^2\mathcal{A}^2\omega^2 l$.

\subsubsection{Green's function method}

As another approach, we can start with a well-defined Cauchy problem
\begin{align}
        &(\partial_t^2-\partial_y^2-\partial_z^2+m_a^2)a=\theta(m_a^2-m_\gamma^2)\gamma^{(0)}\Theta(t)\;, \label{ALPEOM} \\
        & a(t=0,y,z)=0 \;,\quad \partial_t a(t=0,y,z) = 0\;,  
\end{align}
where the Heaviside step function $\Theta(t)$ indicates that the sourced system begins to respond after $t=0$. First, we need to derive the corresponding Green's function which solves the equation
\begin{equation}
    (\partial_t^2-\partial_y^2-\partial_z^2+m_a^2)G_R(t,y,z)=\delta(t)\delta(y)\delta(z)\;. 
\end{equation}
The spatial Fourier transform of this Green's function is well-known:
\begin{equation}
    \tilde{G}_R(t,k_y,k_z)=\frac{\sin(\omega_k t)}{\omega_k}\Theta(t)\;,\quad\omega_k=\sqrt{k_y^2+k_z^2+m_a^2}\;. 
\end{equation}
Therefore, 
\begin{equation}
    G_R(t,y,z)=\frac{1}{(2\pi)^2}\int d^2k\,e^{i(k_yy+k_zz)}\frac{\sin(\omega_k t)}{\omega_k}\Theta(t)\;.
\end{equation}
The solution for the original problem can then be constructed as
\begin{equation}\label{a sol by Green's function}
    \begin{aligned}
        a(t,y,z)&=\theta(m_a^2-m_\gamma^2)\int dt'dz'dy'G_R(t,y,z;t',y',z')\gamma^{(0)}(t',y',z')\Theta(t')\\
        &=\mathcal{A}\frac{\theta(m_a^2-m_\gamma^2)}{(2\pi)^2}\int_{-\infty}^\infty dz'e^{i\beta z'}\int_{0}^{l}dy'\sin(k_\perp y')\int_0^{\infty}dt'e^{-i\omega t'}\cdot\\
        &\qquad\qquad\qquad\qquad\int d^2k\,e^{ik_y(y-y')+ik_z(z-z')}\frac{\sin(\omega_k(t-t'))}{\omega_k}\Theta(t-t')\;.
    \end{aligned}
\end{equation}
First, notice that the integral of $z'$ yields $2\pi\delta(k_z-\beta)$. Then integrating the Fourier momentum $k_z$ out, we can replace all $k_z$ by $\beta$. In the next step, we focus on the time integral
\begin{equation}
    \begin{aligned}
        I_t&=\int_0^\infty dt'e^{-i\omega t'}\sin(\omega_k(t-t'))\Theta(t-t')\\
        &=e^{-i\omega t}\int_0^{\infty}dt'e^{i\omega(t-t')}\sin(\omega_k(t-t')) \Theta(t-t')\\
        &=\frac{e^{-i\omega t}}{2i}\int_{-\infty}^t d(t-t')\left(e^{i(\omega_k+\omega)(t-t')}-e^{-i(\omega_k-\omega)(t-t')}\right) \Theta(t-t')\;.
    \end{aligned}
\end{equation}
For observing a steady behavior, let us take $t\rightarrow\infty$. Then the integral reduces to the Fourier transform of the Heaviside function, which reads
\begin{equation}
    \int_{-\infty}^\infty dt\,e^{i\omega t}\Theta(t)=\frac{i}{\omega}+\pi\delta(\omega)\;.
\end{equation}
Note that in our case, $\omega_k=\sqrt{k_y^2+\beta^2+m_a^2}>\omega$, i.e., $\omega_k\neq\pm \omega$. We simply ignore the delta function part and obtain
\begin{equation}
    I_t=\frac{e^{-i\omega t}}{2i}\left(\frac{i}{\omega_k+\omega}+\frac{i}{\omega_k-\omega}\right)=e^{-i\omega t}\frac{\omega_k}{\omega_k^2-\omega^2}\;.
\end{equation}
Plugging into the solution, we have
\begin{equation}
    a(t,y,z)=\frac{\theta(m_a^2-m_\gamma^2)}{2\pi}e^{-i\omega t+i\beta z}\int_0^l dy'\sin(k_\perp y')\int_{-\infty}^{\infty}dk_y\frac{e^{ik_y(y-y')}}{\omega_k^2-\omega^2}\;.
\end{equation}
Recalling that \eq\eqref{musquare} gives $\omega^2=\mu^2+\beta^2+m_a^2$, we see that the poles of the last integrand are $k_y=\pm i|\mu|$. As before, the fact that $\mu^2<0$ plays an important role in determining the poles with a purely imaginary structure, which leads to an exponential decay (see below). For our interests in the outside regions, for instance, $y<0$, we can pick the contour in the lower half plane: 
\begin{equation}
    \begin{aligned}
        a(t,y,z)&=\mathcal{A}\frac{\theta(m_a^2-m_\gamma^2)}{2\pi} \ex^{-i\omega t+i\beta z}\int_0^l dy' \sin(k_\perp y') (-2\pi i)\frac{\ex^{|\mu|(y-y')}}{-2i|\mu|}\\
        &=\mathcal{A}\theta \frac{k_\perp}{2|\mu|}\left(\ex^{|\mu| y}+e^{|\mu| (y-l)}\right)\ex^{-i\omega t+i\beta z}\;,
    \end{aligned}
\end{equation}
This matches the transverse profile shown in \eq\eqref{fa}. In terms of the split \eqref{ALPSplit} of the ALP field, we see that the free solution $a_{\text{free}}$ vanishes for large times and only the particular solution $a_\gamma$ survives.

As an additional remark, this method also allows us to consider -- instead of \eq \eqref{photonConfined} -- more general localized photon profiles: 
\begin{equation}\label{photonConfinedGeneral}
	\gamma^{(0)} = \mathcal{A} Y(y) \ex^{-i\omega t + i\beta z}\;,
\end{equation}
where now $Y(y)$ no longer is restricted to be $\sin\left(k_\perp y\right) \mathbf{1}_{\{0\leq y\leq l\}}$. Assuming right away that the ALP matches the frequency and longitudinal momentum of the photon solution (\cf \eqs \eqref{ALPSplit} and \eqref{aGamma}),
\begin{equation}
a	= \theta \mathcal{A}\ex^{-i\omega t + i\beta z} f_a(y) \;,
\end{equation}
we derive from the equation of motion \eqref{ALPEOM}:
\begin{equation}
	\left(-\partial_y^2 - \mu^2\right) f_a(y) = (m_a^2 - m_\gamma^2) Y(y) \;.
\end{equation}
This is solved by 
\begin{equation} \label{faGeneral}
	f_a(y) = \int \frac{\diff k_y}{2 \pi} \ex^{i k_y y} \frac{m_a^2 - m_\gamma^2}{k_y^2 - \mu^2} \int \diff y'  \ex^{-i k_y y'} Y(y') = \int \diff y' \frac{m_a^2 - m_\gamma^2}{2|\mu|} \ex^{-|\mu||y - y'|} Y(y')\;,
\end{equation}
which matches \eq \eqref{fa} for $Y(y) = \sin\left(k_\perp y\right) \mathbf{1}_{\{0\leq y\leq l\}}$. We conclude that the decay $f_a(y) \sim \ex^{-|\mu||y|}$ outside the fiber is generic for all compact photon profiles, as long as $\mu^2 < 0$.

\subsubsection{Back-reaction to the photon field}

Plugging the particular solution $a_\gamma$ of the ALP field back into the photon equation, we can analyze the back-reaction up to order $\theta^2$. Namely, we try to find the solution to
\begin{equation}
    (n^2\partial_t^2-\partial_y^2-\partial_z^2)\gamma^{(1)}=gB\omega \theta a_\gamma\equiv\theta^2(m_a^2-m_\gamma^2)a_\gamma\;,
\end{equation}
where $\gamma^{(1)}$ should fulfill the boundary conditions
\begin{equation}
    \gamma^{(1)}(t,y=0,z)=\gamma^{(1)}(t,y=l,z)=0\;.
\end{equation}
The standard approach is to expand the photon field and the source in terms of the orthogonal basis given by $\{\sin(k_m y)\}_{{m\in\mathbb{N}}}$: 
\begin{equation}
    \left( n^{2}\partial _{t}^{2}+k_{m}^{2}+\beta ^{2}\right) \gamma^{(1)}_m\left( t\right) =\theta ^{2}\left( m_{a}^{2}-m_{\gamma }^{2}\right)\mathcal{A} \left[\frac{2}{l}\int_{0}^{l}f_{a}\left( y\right) \sin (k_{m}y) \diff y\right] \cdot \ex^{-i\omega t}\;,
\end{equation}
where the photon mode coefficients $\gamma^{(1)}_m$ and frequencies $k_m$ are defined by
\begin{equation}
    \gamma^{(1)}(t,y,z)=e^{i\beta z}\sum^{\infty }_{m=1}\gamma^{(1)}_m\left( t\right) \sin (k_{m}y)\mathbf{1}_{\{0\leq y\leq l\}}\;,\quad k_{m}=\dfrac{m\pi }{l}\ ({m\in \mathbb{N}})\;.
\end{equation}
The above equation describes a standard-driven oscillator. Only when 
\begin{equation}
    k_m^2+\beta^2=n^2\omega^2\qquad \Leftrightarrow \qquad  k_m=k_1\equiv k_\perp\;, 
\end{equation}
the resonant driving occurs, which implies the accumulative effect in the phase shift. Indeed, a closed form of one particular solution is given by 
\begin{equation}
  \gamma^{(1)}(t,y,z) =   \ex^{-i\omega t+i\beta z} \mathcal{A} \theta^2 \left(h_{\text{res}}(y) +  h_{\text{osc}}(y) \right)\mathbf{1}_{\{0\leq y\leq l\}} \;,
\end{equation}
with
\begin{equation}
	 h_{\text{res}}(y) = t\cdot i \frac{(m_a^2-m_\gamma^2)}{2 n^2 \omega} \left(1 + \frac{2 k_\perp^2\left(1 + \ex^{-|\mu|l}\right)}{|\mu|l\left(m_a^2-m_\gamma^2\right)}\right)\sin\left(k_\perp y\right) \;,
\end{equation}
and
\begin{equation}
    h_{\text{osc}}(y) = \frac{k_\perp}{2 |\mu|}\left[- \left(\ex^{-|\mu| y}+\ex^{|\mu| (y-l)}\right)
	 + \left(1 + \ex^{-|\mu| l}\right) \left(1 -  \frac{2y}{l} \right)\cos\left(k_\perp y\right)\right] \;.
\end{equation}
Importantly, the solution contains a resonance growth in the mode $k_1 = k_\perp$:
\begin{equation}
    \gamma^{(1)}\supset i\mathcal{A}\theta^2\left(\frac{m_a^2-m_\gamma^2}{2n\omega}\frac{t}{n}\right)\sin(k_\perp y)e^{-i\omega t+ i\beta z}\mathbf{1}_{\{0\leq y\leq l\}}\;.
\end{equation}
After we equate the time-dependent term with the propagation distance via the relation $L=t/n$ in medium, the result matches the phase shift shown in \eq\eqref{masterFormulaPhase}. In addition, although the amplitude change is less significant, we point out that its behavior is different from the plane wave result. Namely, the oscillation term $h_\text{osc}(y=l/2)$ implies
\begin{equation}
    \rho_\gamma=\theta^2\frac{k_\perp}{\sqrt{m_a^2+\beta^2-\omega^2}}\ex^{-\sqrt{m_a^2+\beta^2-\omega^2}\frac{l}{2}}\;, 
\end{equation}
which decays faster than $\rho_\gamma\sim2\theta^2$ shown in \eq\eqref{resultMedium} as $m_a$ increases.

\subsection{Bent waveguides}\label{Bent waveguides}

We now consider a bent waveguide with a large bending radius, $R$, placed in an external magnetic field perpendicular to the bending plane. As in the straight case, we pick the photon mode which is uniform along the direction of the magnetic field. In the polar coordinates $(r,\varphi)$, the full system to be solved is
\begin{equation}\label{eomBent}
	\left[ \nabla^2_{r,\varphi} + \omega^2+\begin{pmatrix}
		-m_\gamma^2 & gB \omega \\
		gB \omega & -m_a^2
	\end{pmatrix}\right]\begin{pmatrix}
		{\gamma}(r,\varphi) \\ {a}(r,\varphi)
	\end{pmatrix}=0\;.
\end{equation}
Again, we have assumed the harmonic dependence ${\gamma},{a}\sim \ex^{-i\omega t}$. First, we perturbatively solve the background photon field, which fulfills the equation of motion
\begin{equation}
    \left(\frac{\partial^2}{\partial r^2} + \frac{1}{r} \frac{\partial}{\partial r} + \frac{1}{r^2} \frac{\partial^2}{\partial \varphi^2}+n^2\omega^2\right)\gamma^{(0)}(r,\varphi)=0\;,
\end{equation}
with the boundary conditions
\begin{equation}
    \gamma^{(0)}(r=R,\varphi)=\gamma^{(0)}(r=R+l,\varphi)=0\;.
\end{equation}
The solution is separable, the $r$-dependent part of which should be the combination of Bessel functions of the first and second kind: 
\begin{equation}
    \gamma^{(0)}=\left(c_1J_\nu(n\omega r)+c_2Y_\nu(n\omega r)\right)\ex^{i\nu \varphi
    }\;,
\end{equation}
where the coefficients $c_1$ and $c_2$ are determined by the boundary conditions. While solving $c_1,\,c_2$, the order of the Bessel functions $\nu$ is constrained by
\begin{equation}
    \begin{vmatrix}
        J_{\nu }\left( n\omega R\right) & Y_{\nu }\left( n\omega R\right)  \\
        J_{\nu }\left( n\omega(R+l)\right) & Y_\nu\left( n\omega (R+l)\right) 
    \end{vmatrix}=0\;.
\end{equation}
Solving the above equation analytically is challenging. However, as a leading order approximation, we can estimate the order $\nu$ by comparing with the straight-case result 
\begin{equation}
    \frac{\nu}{R}\approx \beta = \sqrt{n^2\omega^2-k_\perp^2}\;,\qquad k_\perp\equiv\frac{\pi}{l}\;.  
\end{equation}
In fact, a more precise relation, $\nu=\beta R(1+O(l/R))$, can be derived via perturbation, shown in Appendix \ref{app:photonInBentWaveguide}. In the following, it is sufficient to keep only the leading-order term. We learn from the relation $\nu\approx \beta R$ that the order $\nu\sim O(10^6)$ is a very large number for the typical parameters $\beta\sim O(1\text{eV})$ and $R\sim 0.2\text{m}$.

 Next, regarding the photon field as the driving source of the ALP field, we have the equation of motion 
\begin{equation}
    \left(\frac{\partial^2}{\partial r^2} + \frac{1}{r} \frac{\partial}{\partial r} + \frac{1}{r^2} \frac{\partial^2}{\partial \varphi^2}+\omega^2-m_a^2\right)a(r,\varphi)=\begin{cases}
            \begin{aligned}
                &-gB\omega \gamma^{(0)}\;, &&R\leq r\leq R+l\;,  \\
                &0\;, &&\text{otherwise}\;.
            \end{aligned}    
        \end{cases} 
\end{equation}
As in the straight case, the particular solution copies the longitudinal momentum of the photon field, satisfying $a(r,\varphi)=\theta a(r) \ex^{i\nu \varphi}$. We are interested in the behavior outside the waveguide, where the ALP field reads
\begin{equation}
    a(r)=\begin{cases}
    \begin{aligned}
        &c_{\text{decay}} \mathcal{A} J_\nu(\sqrt{\omega^2-m_a^2}\ r)\;, &&r<R\\ 
        &c_\text{rad} \mathcal{A} H^{(1)}_\nu(\sqrt{\omega^2-m_a^2}\ r)\;, &&r>R+l 
    \end{aligned}
    \end{cases} \;. 
\end{equation}
In principle, the solution to the free Bessel equation outside the waveguide should be some combination of the Bessel functions $J_\nu$ and $Y_\nu$. For $r<R$, we pick the first kind since $Y_\nu \rightarrow -\infty$ as the radius goes to 0. Here, the argument $\sqrt{\omega^2-m_a^2}\ r\lesssim n\omega r<\nu$ implies an exponential decay when $r$ decreases.\footnote
{We will see this decay in the large-order approximation of $J_\nu$ below (\cf \eq \eqref{Approx Bessel for large order}).}
For $r>R+l$, we have to use the Hankel function of the first kind $H_\nu^{(1)}\equiv J_\nu + i Y_\nu$, because $H_\nu^{(1)}\ex^{-i\omega t}$ represents an outgoing cylindrical wave. Neither $J_\nu$ nor $Y_\nu$ alone can correctly describe the behavior in this region since both of them contain $H_\nu^{(2)}$, representing the incoming cylindrical wave, which is not allowed. The appearance of the outgoing wave, unlike the straight case, requires a more detailed analysis of this ALP leakage. The loss of ALPs here is analogous to the bending loss of optical fibers, which has been well studied (see \cite{6769759, 1128318, Marcuse:76}, for example). In the following, we will refer to the main idea in \cite{Marcuse1972}. 

The loss amplitude is encoded in the coefficient $c_\text{rad}$. We can approximate it up to the leading order. Considering the asymptotic expansion of the Bessel functions for a large order \cite{WatsonBessel, Gradshteyn:2007}, 
\begin{equation}\label{Approx Bessel for large order}
    \begin{aligned}
        J_\nu(\frac{\nu}{\cosh\alpha})&\approx\frac{1}{\sqrt{2\pi\nu\tanh\alpha}}\ex^{-\nu(\alpha-\tanh\alpha)}\;,\\
        Y_\nu(\frac{\nu}{\cosh\alpha})&\approx-\frac{1}{\sqrt{\frac{\pi}{2}\nu\tanh\alpha}}\ex^{\nu(\alpha-\tanh\alpha)}\;, 
    \end{aligned}
\end{equation}
we can match the parameter $\alpha$ to our argument near the waveguide as
\begin{equation}
    \cosh\alpha\equiv\frac{\nu}{\sqrt{\omega^2-m_a^2}\ r}>1\;, \quad\text{for}\; r\sim R \;.
\end{equation}
With the definition $r\equiv R+y$, we can expand those functions of $\alpha$ up to the order $y/R$: 
\begin{equation}
    \begin{aligned}
        \cosh\alpha &= \frac{\nu}{\sqrt{\omega^2-m_a^2}(R+y)}\approx \tilde{n}-\tilde{n}\frac{y}{R}\;,\\
        \alpha &\approx \text{arccosh}(\tilde{n})-\frac{\tilde{n}}{\sqrt{\tilde{n}^2-1}}\frac{y}{R}\;,\\
        \tanh\alpha &\approx \sqrt{1-\frac{1}{\tilde{n}^2}}-\frac{1}{\tilde{n}\sqrt{\tilde{n}^2-1}}\frac{y}{R}\;,
    \end{aligned}
\end{equation}
where
\begin{equation}
    \tilde{n}\equiv\frac{\nu}{\sqrt{\omega^2-m_a^2}\ R}\approx\frac{\beta}{\sqrt{\omega^2-m_a^2}}\sim n\;. 
\end{equation}
Then, we can compute the exponent in the expansions \eq\eqref{Approx Bessel for large order}
\begin{equation}
    \nu(\alpha-\tanh\alpha)\approx \left(\text{arccosh}(\tilde{n})-\sqrt{1-\frac{1}{\tilde{n}^2}}\right)\beta R-\frac{\sqrt{\tilde{n}^2-1}}{\tilde{n}}\beta y\;.
\end{equation}
By substituting the rough value $\tilde{n}\sim n\sim 1.5$, we notice the first term above is around $0.2\beta R\sim O(10^5)$, which implies $J_\nu \ll Y_\nu$ for $y\ll R$. We can also simplify the second term by recalling the definition of the parameter $\mu^2$ in \eq\eqref{musquare} as 
\begin{equation}
    \frac{\sqrt{\tilde{n}^2-1}}{\tilde{n}}\beta y \approx \sqrt{\beta^2-\omega^2+m_a^2}\ y =|\mu| y\;.  
\end{equation}
Therefore, we have the approximation of the Hankel function: 
\begin{equation}
    H_\nu^{(1)}(\sqrt{\omega^2-m_a^2}\ r)\approx iY_\nu(\frac{\nu}{\cosh\alpha})
    \approx -\frac{i}{\sqrt{\frac{\pi}{2}|\mu|R}}\ex^{\left(\text{arccosh}(\tilde{n})-\sqrt{1-\frac{1}{\tilde{n}^2}}\right)\beta R}\ex^{-|\mu|y}\;.
\end{equation}
As expected, the above $y$-dependent feature is the same as the straight result shown in \eq\eqref{fa}. This allows us to determine the coefficient $c_\text{rad}$ by comparing $a(r)$ and $f_a(y)$ for $r=R+y$ slightly larger than $R+l$, namely,  
\begin{equation}
    c_\text{rad}=i\frac{k_\perp}{2|\mu|}(1+\ex^{|\mu|l})\sqrt{\frac{\pi}{2}|\mu|R}\ \ex^{-\left(\text{arccosh}(\tilde{n})-\sqrt{1-\frac{1}{\tilde{n}^2}}\right)\beta R}\;.
\end{equation}
The ALP leakage can be understood by computing the power loss at infinity, since a simple asymptotic expansion for the Hankel function \cite{Gradshteyn:2007} leads to
\begin{equation}
    a(r\rightarrow\infty)\approx \mathcal{A} c_\text{rad}\sqrt{\frac{2}{\pi\sqrt{\omega^2-m_a^2}\ r}}\ex^{i\sqrt{\omega^2-m_a^2}\ r}\ex^{i(-\frac{\pi}{2}\nu-\frac{\pi}{4})}\;.
\end{equation}
The average power flow density along the radial direction is
\begin{equation}
    S_r=-\frac{1}{2}\partial_ta\partial_r a^* \approx \theta^2\mathcal{A}^2\abs{c_\text{rad}}^2\frac{\omega}{\pi r}\;, 
\end{equation}
where we need to use the full ALP field $a(t,r,\varphi)$ with all the parameters. Meanwhile, the average power flow density of the incident photon field in the waveguide reads 
\begin{equation}
    S_\gamma=\frac{1}{2}E_xH_y^*=\frac{1}{2}\frac{\beta}{\omega}(\omega\mathcal{A})^2\sin^2(k_\perp y)\;.
\end{equation}
Finally, the ratio of the energy loss due to the ALP leakage after a traveling distance $L$ to the given initial photon beam energy becomes 
\begin{equation}
    \begin{aligned}
        \eta=\frac{\left(S_r\frac{r}{R}L\right)\big|_{r\rightarrow\infty}}{\int_0^lS_\gamma dy}&=\frac{\theta^2 k_\perp^2}{2 \beta |\mu|}(1+e^{|\mu|l})^2\frac{L}{l}\ex^{-2\left(\text{arccosh}(\tilde{n})-\sqrt{1-\frac{1}{\tilde{n}^2}}\right)\beta R}\\
        &\sim\theta^2\frac{k_\perp^2}{\beta|\mu|}\frac{L}{l}\ex^{-0.4\beta R}\;.
    \end{aligned}
\end{equation}
Because of the tiny value of the exponential function $\sim\exp{(-0.4\times10^6)}$, even though the large distance yields a factor $L/l \sim 10^5\,\text{m}/\mu\text{m}\sim 10^{11}$, the ALP leakage effect is still completely negligible. 

\subsection{Other higher-order effects}

Already in the case of a straight fiber, there are other mechanisms through which ALPs can leave the fiber. While we have worked to leading order in $g$ throughout this paper, in principle additional channels for ALP leakage can open up at higher orders in $g$. Taking into account that in the relevant parameter space $g \lesssim 10^{-5}\, \text{GeV}^{-1}$ and all other energy scales do not exceed the range of $\text{eV}$ (in particular $\omega \sim |m_\gamma| \sim \text{eV}$, $B\lesssim 10^3\, \text{eV}^2$), we see that processes beyond leading order in $g$ are strongly suppressed. 

Finally, we point out that any realistic fiber exhibits photon loss already in the absence of ALPs.  Microscopically, photons are lost through interactions with the fiber material. In the presence of ALPs, such interaction could also change the momentum of produced ALPs in such a way that they can leave the fiber. However, we expect that this effect only leads to a small correction to the photon loss that already occurs without ALPs. 

\section{Discussion}
\label{sec:summary}
\subsection{Essence of the mechanism}

As we have observed, the transverse motion of ALPs can be effectively described by $(\partial_y^2 + \mu^2)a$, with $\mu^2$ defined in \eq \eqref{musquare}.  In the silica fiber with $n>1$, that ALPs cannot leave the fiber is determined by the fact that $\mu^2<0$. Rewriting
\begin{equation} \label{musquare2}
	\mu^2 = k_\perp^2 + m_\gamma^2 - m_a^2 \;,
\end{equation}
we observe that the difference as compared to the plane wave result of \cite{Raffelt:1987im} is the appearance of $k_\perp \sim 1/l$, which encodes the effect of confinement. However, in the normal silica fiber with large $n>1$, we have $|m_\gamma^2| \gg m_a^2$ and $|m_\gamma^2| \gg k_\perp^2$. Since $k_\perp^2$ can be neglected, the effect of confinement only leads to a small correction and the plane-wave result of \cite{Raffelt:1987im} remains a good approximation, as we have shown.

In our calculations, we have seen that ALPs could leave the fiber for $\mu^2>0$. Heuristically, this can be explained as follows: A photon of energy $\omega$ and momentum $\beta$ converts to an ALP with the same energy and longitudinal momentum. Thus, the ALP has the dispersion $0=\omega^2 - \beta^2 -m_a^2 -k_{\perp,a}^2 = 	\mu^2 -k_{\perp,a}^2$, where $k_{\perp,a}$ is the ALP transverse momentum. In contrast to the photon, the ALP is not confined and so there is no lower bound on $k_{\perp,a}$. As a result, the solution can select an appropriate $k_{\perp,a}$ corresponding to an on-shell ALP for all $\mu^2 >0$ -- see appendix \ref{app:HamiltonianMixing} for a more detailed argument.

The value $\mu^2=0$, which separates the two regimes discussed above, has a special role as this is the condition for resonant conversion of photons into ALPs (\cf \cite{Raffelt:1987im}). In summary, we expect that \eq \eqref{musquare2} distinguishes three cases:
\begin{itemize}
	\item[(i)] \emph{Confinement}: $\mu^2<0$. When photons oscillate into ALPs, the ALPs cannot propagate outside the fiber. The photon field does not lose energy.
	\item[(ii)] \emph{Resonance}:  $\mu^2=0$. Photons convert resonantly into ALPs and a large fraction of the photon energy can be lost.
	\item[(iii)] \emph{Leakage}: $\mu^2>0$. Photons oscillate into ALPs, which can leave the fiber. A non-zero but generically small fraction of the photon energy is lost.
\end{itemize}

\subsection{Outlook: hollow-core fiber}
\label{ssec:hollowCore}
As discussed,  the normal silica fiber always realizes case (i), \ie confinement, since $n>1$ implies that $m_\gamma^2$ is negative and that $|m_\gamma^2|$ is larger than all other scales. It becomes immediately clear that the situation is distinctly different in a hollow-core fiber, where the effective refractive index can be close to unity, $n\approx 1$. In this case, we expect leakage to occur for sufficiently small $m_a$. Moreover, we can achieve resonance for $m_a\sim k_\perp \sim 1/l$, \ie when the ALP mass is on the order of the inverse fiber radius. This has two immediate implications. 
\begin{enumerate}
	\item  Hollow-core fibers are significantly better-suited for ALP searches than silica fibers. Leakage and especially resonance enhance the signal, in particular since now ALPs can be emitted easily and so the photon amplitude decreases.
	\item If $m_\gamma^2$ is small, $k_\perp$ plays an important role and the plane-wave result of \cite{Raffelt:1987im} no longer represents a good approximation. Among others, resonance occurs for smaller values of $m_\gamma^2$ (equivalently larger values of $m_a^2$) as compared to a situation without transverse boundary conditions.
\end{enumerate}

Because of the ability to achieve resonance, the use of straight hollow-core fibers for detecting light weakly-interacting particles such as ALPs has already been proposed in the WISPFI experiment \cite{Batllori:2023gwy,Batllori:2025ogo}. However, the formulas of \cite{Raffelt:1987im}, which do not take into account the effect of transverse boundary conditions, were employed, and so the forecasts of \cite{Batllori:2023gwy,Batllori:2025ogo} will need to be revisited.

\section{Conclusion}
\label{sec:conclusion}

The leading method for axion-like particle (ALP) detection is through their mixing with photons in an external magnetic field. The observable effects of this interaction generically grow with the distance traveled by the photon, making it desirable to achieve long propagation lengths. To this end, we have proposed a new detection strategy based on long, coiled optical fibers, which naturally allow for propagation lengths of order $10^{5}\,\text{m}$ while still permitting a spatially compact experimental realization.

The primary goal of this work was to establish the theoretical framework required to analyze photon--ALP conversion in such systems. Compared to the conventional plane-wave treatment, two additional ingredients must be taken into account when photons propagate in optical fibers: the transverse boundary conditions imposed by the fiber geometry and the effects induced by bending. We have developed a systematic description that incorporates both effects and allows for a controlled assessment of their relevance.

Focusing on solid silica fibers with refractive index significantly larger than unity, we have shown that the effects of both transverse confinement and bending are suppressed. As a consequence, photon loss due to conversion into ALPs is negligible, and ALPs produced through mixing remain effectively confined within the fiber. In this regime, the leading observable signal is a phase shift of the photon mode, analogous to the plane-wave case. We show the resulting sensitivity to the photon--ALP coupling in \eqref{masterFormulaCoupling}, which highlights the improvement with fiber length. For suitable phase sensitivities, this setup can probe ALPs in the large-mass region that is poorly constrained by existing laboratory experiments.

Beyond this specific realization, we have identified the conditions under which transverse boundary effects and photon loss become important. In particular, we have shown that these effects play a central role in hollow-core fibers, where the effective refractive index is close to (and slightly smaller than) unity. In this case, transverse confinement strongly modifies photon--ALP mixing, leading to substantial deviations from the plane-wave result and enabling resonant conversion into ALPs. Consequently, hollow-core fibers can yield significantly stronger signals and provide access to a much broader region of parameter space.

We leave a quantitative analysis of ALP detection with hollow-core fibers for an upcoming paper \cite{ALP2}. Taken together, the results presented here demonstrate that long, bent optical fibers -- especially hollow-core fibers -- provide a promising new approach for ALP searches.

	\acknowledgments 
We thank the GRAVITES collaboration for useful discussions and especially Christopher Hilweg for valuable feedback on the manuscript. This work was supported in part by the Humboldt Foundation under Humboldt Professorship Award, by the European Research Council Gravities Horizon Grant AO number: 850 173-6,
by the Deutsche Forschungsgemeinschaft (DFG, German Research Foundation) under Germany's Excellence Strategy - EXC-2111 - 390814868, and Germany's Excellence Strategy under Excellence Cluster Origins. All authors were supported by the European Research Council Gravites Horizon Grant AO number: 850 173-6. 

\paragraph*{Disclaimer} Funded by the European Union ERC, GRAVITES, 101071779. Views and opinions expressed are however those of the author(s) only and do not necessarily reflect those of the European Union or the European Research Council. Neither the European Union nor the granting authority can be held responsible for them.

\appendix

\section{General solution to the mixing system}\label{app:generalSolution}

The most general solution to the diagonalized system \eqref{eomGeneralDiagonal} reads
\begin{equation}
    \begin{pmatrix}
        \gamma'(z) \\ a'(z)
    \end{pmatrix}=\begin{pmatrix}
        A_1 e^{ik_+'z}+B_1e^{-ik_+'z} \\ A_2 e^{ik_-'z}+B_2e^{-ik_-'z} 
    \end{pmatrix}\;,
\end{equation}
where $A_{1,2},\ B_{1,2}$ are undetermined coefficients. Rotated to the original fields, it gives the solution to the photon and ALP
\begin{equation}
    \begin{pmatrix}
        \gamma(z) \\ a(z)
    \end{pmatrix}=V^T\begin{pmatrix}
        \gamma'(z) \\ a'(z)
    \end{pmatrix}=\begin{pmatrix}
        \cos\theta(A_1e^{ik_+'z}+B_1e^{-ik_+'z})-\sin\theta (A_2 e^{ik_-'z}+B_2e^{-ik_-'z}) \\
        \sin\theta(A_1e^{ik_+'z}+B_1e^{-ik_+'z})+\cos\theta (A_2 e^{ik_-'z}+B_2e^{-ik_-'z})
    \end{pmatrix}\;.
\end{equation}
After imposing complete initial conditions 
\begin{equation}
    \begin{pmatrix}
        \gamma(z) \\ a(z)
    \end{pmatrix}\biggr|_{z=0}=\begin{pmatrix}
        \mathcal{A}_\gamma \\ \mathcal{A}_a
    \end{pmatrix}\;, \quad
    \begin{pmatrix}
        \partial_z\gamma(z) \\ \partial_z a(z)
    \end{pmatrix}\biggr|_{z=0}=\begin{pmatrix}
        \mathcal{A}'_\gamma \\ \mathcal{A}'_a
    \end{pmatrix}\;, 
\end{equation}
we can fix all the undetermined coefficients as 
\begin{equation}
    \begin{aligned}
        A_1&=\frac{1}{2}\left(\cos\theta \mathcal{A}_\gamma+\sin\theta \mathcal{A}_a+\frac{\cos\theta \mathcal{A}'_\gamma+\sin\theta \mathcal{A}'_a}{ik_+'}\right)\;,\\
        B_1&=\frac{1}{2}\left(\cos\theta \mathcal{A}_\gamma+\sin\theta \mathcal{A}_a-\frac{\cos\theta \mathcal{A}'_\gamma+\sin\theta \mathcal{A}'_a}{ik_+'}\right)\;,\\
        A_2&=\frac{1}{2}\left(-\sin\theta \mathcal{A}_\gamma+\cos\theta \mathcal{A}_a+\frac{-\sin\theta \mathcal{A}'_\gamma+\cos\theta \mathcal{A}'_a}{ik_-'}\right)\;,\\
        B_2&=\frac{1}{2}\left(-\sin\theta \mathcal{A}_\gamma+\cos\theta \mathcal{A}_a-\frac{-\sin\theta \mathcal{A}'_\gamma+\cos\theta \mathcal{A}'_a}{ik_-'}\right)\;. 
    \end{aligned}
\end{equation}
To simplify the notations, we define the free longitudinal momenta
\begin{equation}
    \beta_\gamma\equiv\sqrt{\omega^2-m_\gamma^2}\;, \quad\beta_a\equiv\sqrt{\omega^2-m_a^2}\;,
\end{equation}
so that we get to the leading order of $\theta^2$, 
\begin{equation}
    \begin{aligned}
        e^{\pm ik_+'z}&=e^{\pm i\beta_\gamma z}\left(1 \pm i\theta^2\frac{\Delta m^2}{2\beta_\gamma}z\right)\;,\\
        e^{\pm ik_-'z}&=e^{\pm i\beta_a z}\left(1\mp i\theta^2\frac{\Delta m^2}{2\beta_a}z\right)\;,
    \end{aligned}
\end{equation}
where $\Delta m^2\equiv m_a^2-m_\gamma^2$. 

For the ideal pure-photon incidence, 
\begin{equation}
    \mathcal{A}_\gamma=\mathcal{A}\;, \quad \mathcal{A}_a=0\;, \quad\mathcal{A}_\gamma'=i\beta_\gamma \mathcal{A}\;, \quad \mathcal{A}_a'=0\;, 
\end{equation}
we have
\begin{equation}
    \begin{aligned}
        A_1&=\mathcal{A}\left[1-\frac{\theta^2}{2}\left(1+\frac{\Delta m^2}{2\beta_\gamma^2}\right)\right]\;, \quad B_1=\mathcal{A}\frac{\theta^2}{2}\left(\frac{\Delta m^2}{2\beta_\gamma^2}\right)\;\;,\\
        A_2&=-\mathcal{A}\frac{\theta}{2}\left(1+\frac{\beta_\gamma}{\beta_a}\right)\;, \quad B_2=-\mathcal{A}\frac{\theta}{2}\left(1-\frac{\beta_\gamma}{\beta_a}\right)\;.
    \end{aligned}
\end{equation}
The final solution for the photon field is
\begin{equation}
    \begin{aligned}
        \gamma(z)=\mathcal{A}\left(1-\rho_\gamma+i\phi_\gamma\right)e^{i\beta_\gamma z}+\mathcal{A}\frac{\theta^2}{2}\left[\frac{\Delta m^2}{2\beta_\gamma^2}+\left(1-\frac{\beta_\gamma}{\beta_a}\right)e^{i\Delta \omega z}\right]e^{-i\beta_\gamma z}\;,
    \end{aligned}
\end{equation}
where
\begin{equation}
    \begin{aligned}
        \rho_\gamma&=\theta^2\left(2\sin^2\left(\frac{\Delta \omega z}{2}\right)+\frac{1}{2}\left(\left(1-\frac{\beta_\gamma}{\beta_a}\right)\cos(\Delta\omega z)+\frac{\Delta m^2}{2\beta_\gamma^2}\right)\right)\;,\\
        \phi_\gamma&=\theta^2\left(\frac{\Delta m^2}{2\beta_\gamma}z-\sin(\Delta\omega z)+\frac{1}{2}\left(1-\frac{\beta_\gamma}{\beta_a}\right)\sin(\Delta\omega z)\right)\;, 
    \end{aligned}
\end{equation}
and as in the main part we have defined $\Delta \omega = \sqrt{\omega^2 - m_\gamma^2} - \sqrt{\omega^2 - m_a^2}=\beta_\gamma-\beta_a$.

Only in the limit $\beta_\gamma^2\approx \beta_a^2\gg \Delta m^2$ such that $A_1\approx \mathcal{A}\cos\theta,\, A_2\approx-\mathcal{A}\sin\theta,\,B_{1,2}\approx0$, the solution reduces to \eq \eqref{solGeneral}. Practically, this happens for small-mass ALPs and $m_\gamma=0$ in vacuum. Correspondingly, the final solution matches \eq \eqref{resultVacuum}. On the contrary, as long as $|m_\gamma^2|\gg m_a^2$ in media such that $\Delta m^2\sim\beta_\gamma^2\gg \beta^2_a$, the modification for $\rho_\gamma$, $\phi_\gamma$, as well as the backward-moving mode cannot be neglected. Importantly, however, the distance-accumulative effect in the phase shift is still preserved.\footnote{All other terms with $z$ in the oscillating function already have $\theta^2$ as a prefactor, such that they are bounded by $\theta^2$ rather than proportional to $\theta^2 z$.}

\section{Photon field solution in bent waveguide}\label{app:photonInBentWaveguide}

Here, we are solving the photon field in the bent waveguide with a large radius $R$ to the subleading order using perturbation theory. The equation of motion reads
\begin{equation}
    (\nabla^2_{r,\varphi}+n^2\omega^2)\gamma(r,\varphi)=0\;.
\end{equation}
In polar coordinates, the 2-dimensional Laplace operator can be expanded as
\begin{equation}
    \nabla^2_{r,\varphi}=\frac{\partial^2}{\partial r^2} + \frac{1}{r} \frac{\partial}{\partial r} + \frac{1}{r^2} \frac{\partial^2}{\partial \varphi^2}\approx \frac{\partial^2}{\partial y^2} + \frac{1}{R} \frac{\partial}{\partial y} + \left(1-\frac{2y}{R}\right) \frac{\partial^2}{\partial z^2}\;,
\end{equation}
where we defined
\begin{align}
	z & \equiv R \varphi \;,\\
	y & \equiv r-R \;.
\end{align}
In these coordinates, the boundary conditions are still given by
\begin{equation}
    \gamma(y=0,z)=\gamma(y=l,z)=0\;.
\end{equation}
The leading order solution is nothing different from the straight case, \ie
\begin{equation}
    \gamma^{(0)}(y,z)=\mathcal{A}\sin\left(k_\perp y\right) \ex^{i\beta z}\;,\quad\beta=\sqrt{n^2\omega^2 - k_\perp^2}\;,\quad k_\perp \equiv\frac{\pi}{l} \;.
\end{equation}
When the first subleading order is included, the solution becomes 
\begin{equation}
	{\gamma}^{(1)}_\text{bending}(y,z) = \mathcal{A}\ex^{i (\beta + \delta \beta) z} \left( \sin\left(k_\perp y\right) + h_\gamma(y)\right) \;,
\end{equation}
where $\delta \beta$ and $h_\gamma(y)$ are determined by 
\begin{equation}
	\left( \frac{\partial^2}{\partial y^2} + k_\perp^2 \right) h_\gamma(y) = \left(-\frac{1}{R} \frac{\partial}{\partial y} - \frac{2y}{R} \beta^2 + 2 \beta \delta \beta \right) \sin\left(k_\perp y\right) \;.
\end{equation}
It is straightforward to find an explicit solution
\begin{equation}
	\delta \beta = \frac{{l}}{R}\frac{\beta}{2} \;,
\end{equation}
\begin{equation}
    h_\gamma(y) =  -\frac{{y}}{2R}\left[\sin\left({k_\perp y}\right) + \frac{\beta^2}{k_\perp^2} \big(\sin(k_\perp y)+k_\perp (l-y) \cos\left(k_\perp y\right)\big)\right] \;. 
\end{equation}
Translated back to the quantities in the polar coordinates, it shows that the order of the Bessel function we use in section \ref{Bent waveguides} takes the relation
\begin{equation}
    \nu=\beta R\left[1+\frac{1}{2}\frac{l}{R}+o\left(\frac{l^2}{R^2}\right)\right]\;. 
\end{equation}

\section{Hamiltonian argument}
\label{app:HamiltonianMixing}
We can understand our findings in terms of the effective two-dimensional Hamiltonian
\begin{equation}
\begin{aligned}
    &\hat{H} = \int \diff y \diff z\  \frac{1}{2} \left( n^2(\partial_t \hat{\gamma})^2 + ({\nabla} \hat{\gamma})^2 + (\partial_t \hat{a})^2 + ({\nabla} \hat{a})^2 +m_a^2 \hat{a}^2 \right) + \tilde{g}\hat{V}_\text{mix}\;, \\
    &\hat{V}_\text{mix} \equiv  \int \diff y \diff z\ {\hat{a} \hat{\gamma} \;, }
\end{aligned}
\end{equation}
where $\hat{\gamma}$ and $\hat{a}$ represent the photon and ALP field, respectively, and $\tilde{g} \sim g$ is an effective coupling. 

The previous dynamics given by Green's function can be explicitly reproduced in the quantum language. In the following, we work in the interaction picture. The initial state {$\ket{i}\equiv\ket{i}_\gamma\otimes\ket{0}_a$} is chosen such that
\begin{equation}
    \bra{i}\hat{\gamma}(t,\mathbf{x})\ket{i}= \gamma^{(0)}(t,y,z)\;,\quad\bra{i}\hat{a}(t,\mathbf{x})\ket{i}=0\;, 
\end{equation}
where $\mathbf{x}$ denotes the coordinates $(y,z)$. This can be achieved if $\ket{i}_\gamma$ is a coherent state.
The quantum evolution yields the final state $\ket{f}$. Namely, to the leading order, 
\begin{equation}\label{quantum evolution}
    \ket{f} =\hat{T}\exp\left\{-i\tilde{g}\int_0^t \diff t' \hat{V}_\text{mix}(t')\right\}\ket{i}=\left(1-i\tilde{g}\int_0^t \diff t' \hat{V}_\text{mix}(t')+O(\tilde{g}^2)\right)\ket{i}\;,
\end{equation}
which gives the expectation value of the ALP field 
\begin{equation}
\begin{aligned}
    \bra{f}\hat{a}(t,\mathbf{x})\ket{f}&\approx\bra{i}\left(1+i\tilde{g}\int_0^t \diff t' \hat{V}_\text{mix}(t')\right)\hat{a}(t,\mathbf{x})\left(1-i\tilde{g}\int_0^t \diff t' \hat{V}_\text{mix}(t')\right)\ket{i}\\
    &\approx\bra{i}\hat{a}\ket{i}+i\tilde{g}\int_0^t \diff t'\bra{i}\left[\hat{V}_\text{mix}(t'),\hat{a}(t,\mathbf{x})\right]\ket{i}\\
    &=i\tilde{g}\int_0^\infty \diff t'\Theta(t-t')\int \diff^2\mathbf{x}'\ \bra{{i}}\hat{\gamma}(t',\mathbf{x'})\ket{{i}}_\gamma\bra{0}\left[\hat{a}(t',\mathbf{x'}),\hat{a}(t,\mathbf{x})\right]\ket{0}_a\\
    &=\tilde{g}\int_0^\infty \diff t' \int d^2\mathbf{x}'G_R(t,\mathbf{x};t',\mathbf{x}')\gamma^{(0)}(t',\mathbf{x}')\;, 
\end{aligned}
\end{equation}
where the relation between the retarded Green's function and the commutator of ALP fields reads
\begin{equation}
    G_R(t,\mathbf{x};t',\mathbf{x}')\equiv i \Theta(t-t')\bra{0}\left[\hat{a}(t',\mathbf{x'}),\hat{a}(t,\mathbf{x})\right]\ket{0}_a\;.
\end{equation}
{Indeed, we see that $\bra{f}\hat{a}\ket{f}$ yields the Green's function result \eqref{a sol by Green's function}.}

As a complementary argument, we can also derive qualitative features of our ALP solution from mixing with a generic photon state. For the ALP field, we take the usual plane-wave expansion
\begin{equation}
	\hat{a} = \int \frac{\diff^2 \mathbf{q}}{(2 \pi)^2} \frac{1}{\sqrt{2 \varepsilon_\mathbf{q}}}\left( \hat{b}_\mathbf{q} \ex^{-i\varepsilon_\mathbf{q}t+i\mathbf{q}\cdot\mathbf{x}} + \hat{b}_\mathbf{q}^\dagger \ex^{i\varepsilon_\mathbf{q}t-i\mathbf{q}\cdot\mathbf{x}}\right)  \;,
\end{equation}
where
\begin{equation}
\begin{aligned}
    \left[\hat{b}_\mathbf{q}, \hat{b}_{\mathbf{q}'}^\dagger  \right] = (2 \pi)^2 \delta^2(\mathbf{q}-\mathbf{q}') \;&, \quad \varepsilon_\mathbf{q}^2 =\mathbf{q}^2 + m_a^2\;, \\
    \quad \mathbf{q}=(q_y,q_z)\;&,\quad \mathbf{x}=(y,z) \;.
\end{aligned}
\end{equation}
For the photon field, we have the expansion due to the boundary effects in fibers: 
\begin{equation}
	\hat{\gamma} =  \int\frac{\diff k_z}{2\pi}\sum_m \frac{1}{n\sqrt{2\omega_\mathbf{k}l}} \sin\left(k_m y\right) \mathbf{1}_{\{0\leq y\leq l\}}\left(\hat{c}_\mathbf{k}\ex^{-i\omega_\mathbf{k} t + i k_z z} + \hat{c}_\mathbf{k}^\dagger\ex^{i\omega_\mathbf{k} t - i k_z z}\right) \;,
\end{equation}
where
\begin{equation}
\begin{aligned}
    \left[\hat{c}_\mathbf{k},\hat{c}_{\mathbf{k}'}^\dagger\right]=2\pi\delta(k_z-k_z')\delta_{mm'}\;&, \quad \omega _\mathbf{k}^2 = \frac{1}{n^2} \left(k_m^2 + k_z^2\right)\;,  \\
    \quad \mathbf{k}=(k_m,k_z)\;&, \quad k_m=\frac{m\pi}{l}\;.
\end{aligned}
\end{equation}
Thus, we get the Hamiltonian
\begin{equation}
	\hat{H} = \int \frac{\diff k_z}{2\pi} \sum_m \omega_\mathbf{k} \hat{c}_\mathbf{k}^\dagger \hat{c}_\mathbf{k} + \int \frac{\diff^2 \mathbf{q}}{(2 \pi)^2} \varepsilon_\mathbf{q} \hat{b}_{\mathbf{q}}^\dagger \hat{b}_{\mathbf{q}}  +  \tilde{g}\hat{V}_\text{mix} \;, 
\end{equation}
where $\hat{V}_\text{mix}$ denotes the interaction 
\begin{equation}
    \begin{aligned}
        \hat{V}_\text{mix} & = \int \frac{\diff^2 \mathbf{q}}{(2 \pi)^2} \int \frac{\diff k_z}{2\pi}\sum_m \left[ \frac{1}{\sqrt{2\varepsilon_\mathbf{q}}\ n\sqrt{2\omega_\mathbf{k} l}}   \ex^{-i\left( \omega_\mathbf{k} - \varepsilon_\mathbf{q}\right)t}  \hat{b}_\mathbf{q}^\dagger \hat{c}_\mathbf{k} \right.\\
        & \left. \quad \times \int_0^l \diff y \, \sin\left(k_m y\right)  \ex^{-i q_y y}   \int_{-\infty}^\infty \diff z\,  \ex^{i\left(k_z-q_z\right) z} + \left(\text{terms with}\  \hat{b}\hat{c}, \hat{b}^\dagger\hat{c}^\dagger,\hat{c}^\dagger\hat{b}\right) \right] \\
	& = \int \frac{\diff^2 \mathbf{q}}{(2 \pi)^2}  \sum_m \frac{1}{\sqrt{2\varepsilon_\mathbf{q}}\ n\sqrt{2\omega_{k_m,q_z} l}}\ex^{-i\left(\omega_{k_m,q_z} - \varepsilon_\mathbf{q}\right)t} \tilde{f}_{\gamma,m}(q_y) \hat{b}_\mathbf{q}^\dagger \hat{c}_{k_m,q_z} + \cdots\;,
    \end{aligned}
\end{equation}
and we defined 
\begin{equation}
	\tilde{f}_{\gamma,m}(q_y) \equiv \int_0^l \diff y \, \sin\left(k_m y\right)  \ex^{-i q_y y} = \frac{1}{2} \left(\frac{\ex^{-i(q_y - k_m)l}-1}{q_y - k_m} - \frac{\ex^{-i(q_y + k_m)l}-1}{q_y + k_m}\right) \;.
\end{equation}
Here, we only explicitly keep the term describing photon annihilation and ALP creation.

\begin{figure}
	\centering
    \includegraphics[width=0.6\textwidth]{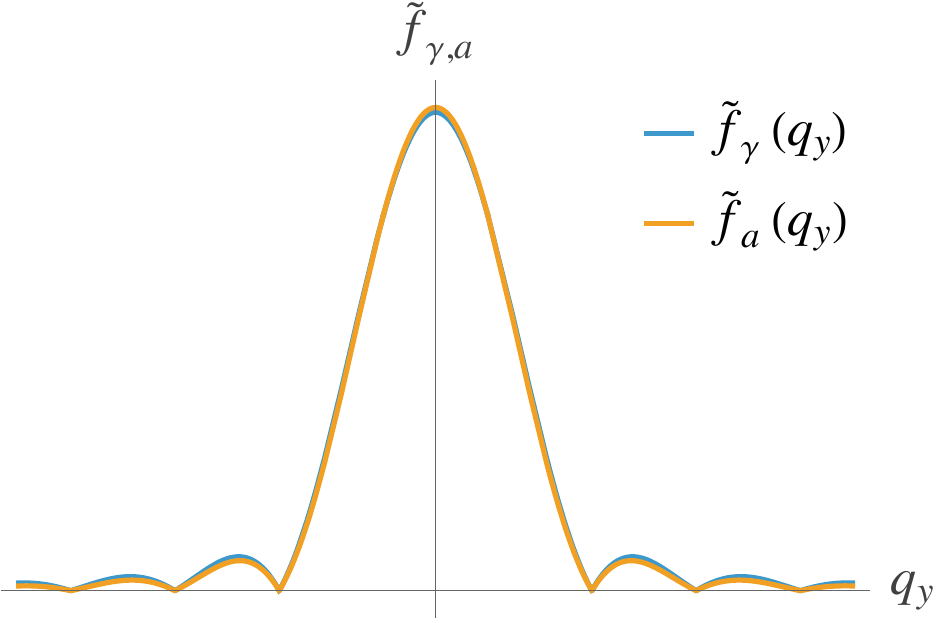}
	\caption{A sketch for the comparison of the photon and ALP transverse profiles in momentum space. Here, $\tilde{f}_{\gamma}(q_y)$ and $\tilde{f}_{a}(q_y)$ are obtained by the Fourier transform of $\sin(k_\perp y)\mathbf{1}_{\{0\leq y\leq l\}}$ and $f_a(y)$, respectively. We use the absolute value for plotting. }
	\label{fig:transverseMomenta}
\end{figure}

It is important to note that translation invariance in $z$-direction implies that the ALP and photon longitudinal momenta must be identical. As for the transverse direction, ${\tilde{f}_{\gamma}(q_y)}\equiv\tilde{f}_{\gamma,m=1}(q_y)$ simply corresponds to the Fourier transform of the transverse profile of the ground photon mode.\footnote
{Equivalently, this behavior can be understood as follows: $\tilde{f}_{\gamma}(q_y)$ is roughly constant for $q_y^2 \lesssim k_\perp^2$. Since the number of modes with which mixing is effective increases the interaction strength (see \cite{Dvali:1999cn}), the photon will mix with all modes $\hat{b}_{q_y, \beta}^\dagger$ in the interval $q_y^2 \lesssim k_\perp^2$.}
This determines which momenta $q_y$ are present for the ALP field, and so also the distribution of transverse momenta of the ALP field is very similar to its counterpart of the photon field -- see comparison in \fig \ref{fig:transverseMomenta}. Thus, the momentum structure of the photon field is copied verbatim to the ALP field, which matches our observation in the main part that the driven ALP solution inherits the structure of the photon source. Of course, the details are determined by dynamics.

For concreteness, we can consider the quantum evolution of a one-photon state with the transverse momentum $k_{m=1}=k_\perp$:
\begin{equation}
    \ket{i}=\ket{1_{m=1,\ k_z=\beta}}_\gamma\otimes\ket{0}_a\;.
\end{equation}
 By convention, the one-particle states are defined as 
\begin{equation}
    \ket{1_{\mathbf{k}}}_\gamma=n\sqrt{2\omega_\mathbf{k}l}\hat{c}_\mathbf{k}^\dagger\ket{0}_\gamma\;,\quad \ket{1_\mathbf{q}}_a=\sqrt{2\varepsilon_\mathbf{q}}\hat{b}_\mathbf{q}^\dagger\ket{0}_a\;. 
\end{equation}
Recalling \eq\eqref{quantum evolution}, we have
\begin{equation}
\begin{aligned}
    \ket{f}&=\left(1-i\tilde{g}\int_0^t dt' \hat{V}_\text{mix}(t')+O(\tilde{g}^2)\right)\ket{i}\\
    &\approx\ket{i}-i\tilde{g}\int\frac{dq_y}{2\pi}\frac{1}{2\varepsilon_{\mathbf{q}}}\int_0^tdt'\ex^{-i(\omega-\varepsilon_{\mathbf{q}})t'}\tilde{f}_\gamma(q_y)\ket{0}_\gamma\otimes\ket{1_{\mathbf{q}}}_a\;,
\end{aligned}
\end{equation}
where
\begin{equation}
    \omega=\frac{1}{n}\sqrt{k_\perp^2+\beta^2}\;,\quad \mathbf{q}=(q_y,\beta)\;
\end{equation}
in the last line. The integral with respect to time yields a function with
\begin{equation}
    \omega-\varepsilon_\mathbf{q}=\omega-\sqrt{q_y^2+\beta^2+m_a^2}
\end{equation}
in the denominator, which implies poles at\footnote
{The square root also gives branch points $q_y=\pm i\sqrt{\beta^2+m_a^2}$, which should be taken into account for contour integrals. However, the contribution from the branch cut does not change the behavior of the poles.}
\begin{equation}
    \omega^2 = m_a^2 + q_y^2 + \beta^2 \qquad \Leftrightarrow \qquad q_y^2 = \mu^2  \;.
\end{equation}

Thus, we see the three cases observed before by computing the wavefunction-like quantity
\begin{equation}
    \bra{0}\hat{a}(t,\mathbf{x})\ket{f}\;,
\end{equation}
in which
\begin{equation}
    \bra{0}\hat{a}(t,\mathbf{x})\ket{1_\mathbf{q}}_a=\ex^{-i\varepsilon_\mathbf{q}t+iq_y y+i\beta z}\;.
\end{equation}
If $\mu^2<0$, the purely imaginary poles $q_y=\pm i|\mu|$ indicate an exponential decay along the $y$ direction after integrating against $q_y$, which corresponds to confinement. For $\mu^2= 0$, the pole becomes the origin, which needs a special treatment -- this is resonance. For large positive $\mu^2$, the poles $q_y=\pm \mu$ lie on the real axis. Taking Feynman's trick that shifts the poles by $\ O(i\varepsilon)$ in the integrand shows the oscillation behavior $\ex^{\pm i\mu y}$, which is the regime of leakage.

\bibliographystyle{JHEP}
\bibliography{references}{}

\end{document}